\documentclass[a4paper,11pt]{article}

\usepackage{jinstpub} 
\usepackage{array,multirow,graphicx}

\usepackage{lineno}

\title{Emittance preservation in advanced accelerators}

\author{Carl A. Lindstr{\o}m}
\author{and Maxence Th{\'e}venet}
\affiliation{Deutsches Elektronen-Synchrotron DESY, Notkestr. 85, 22607 Hamburg, Germany}

\emailAdd{carl.a.lindstroem@desy.de}
\emailAdd{maxence.thevenet@desy.de}

\abstract{Emittance is a beam quality that is vital for many future applications of advanced accelerators, such as compact free-electron lasers and linear colliders. In this paper, we review the challenges of preserving the transverse emittance during acceleration, both inside and outside accelerator stages. Sources of emittance growth range from space charge and instabilities caused by transverse wakefields, which can occur in any advanced accelerator scheme regardless of medium or driver type, to sources more specific to plasma accelerators, such as mismatching, misalignment, ion motion, Coulomb scattering, chromaticity between stages, and more.}


\proceeding{ICFA Beam Dynamics Newsletter, Issue 83}

\begin{document}
\maketitle
\flushbottom

\section{Introduction}

\subsection{Advanced accelerators}

Advanced accelerators aim to reduce the size and cost of producing high-energy particles, using a variety of novel accelerating media and drivers. Most advanced accelerators are based on accelerating particles in the wakefield trailing behind a laser pulse or particle beam, the \textit{driver}, in a medium of either dielectric structures or plasma. Popular combinations include laser-plasma accelerators~\cite{Esarey2009}, beam-driven plasma-wakefield accelerators~\cite{Hogan2016,Adli2016}, beam-driven structure-wakefield accelerators~\cite{Jing2016}, and dielectric laser accelerators~\cite{England2014}. However, by increasing the accelerating gradient, these new techniques also make it more difficult to maintain a high beam quality. This is true both in the longitudinal phase space, resulting in large energy spreads and chirps, as well as in the transverse phase space, resulting in large emittances. Here, we explore the latter---what sources of emittance growth exist in advanced accelerators, and how can we suppress these to preserve the emittance?

This paper consists of two main sections to reflect the basic structure of future accelerator facilities. We will assume that the particle bunch already exists, and will therefore not discuss the generation of low-emittance bunches---we consider only \textit{external injection}, as opposed to \textit{internal injection}. Firstly, in Section~\ref{sec:emittance-growth-in-stage}, we cover the many sources of emittance growth that affect the particle beam inside an accelerating stage, some of which are common to all advanced accelerators and others that are specific to plasma accelerators. Secondly, while most advanced-accelerator experiments to date have focused on single-stage systems, this may not be the case in the future: multiple stages may be required to reach higher energy and/or higher stability~\cite{Lindstrom2021c}. In Section~\ref{sec:emittance-growth-between-stages}, we explore the sources of emittance growth that can result from transport between stages and the process of exchanging the driver.

\subsection{What is emittance?}

\begin{figure}[t]
    \centering
    \includegraphics[width=0.5\linewidth]{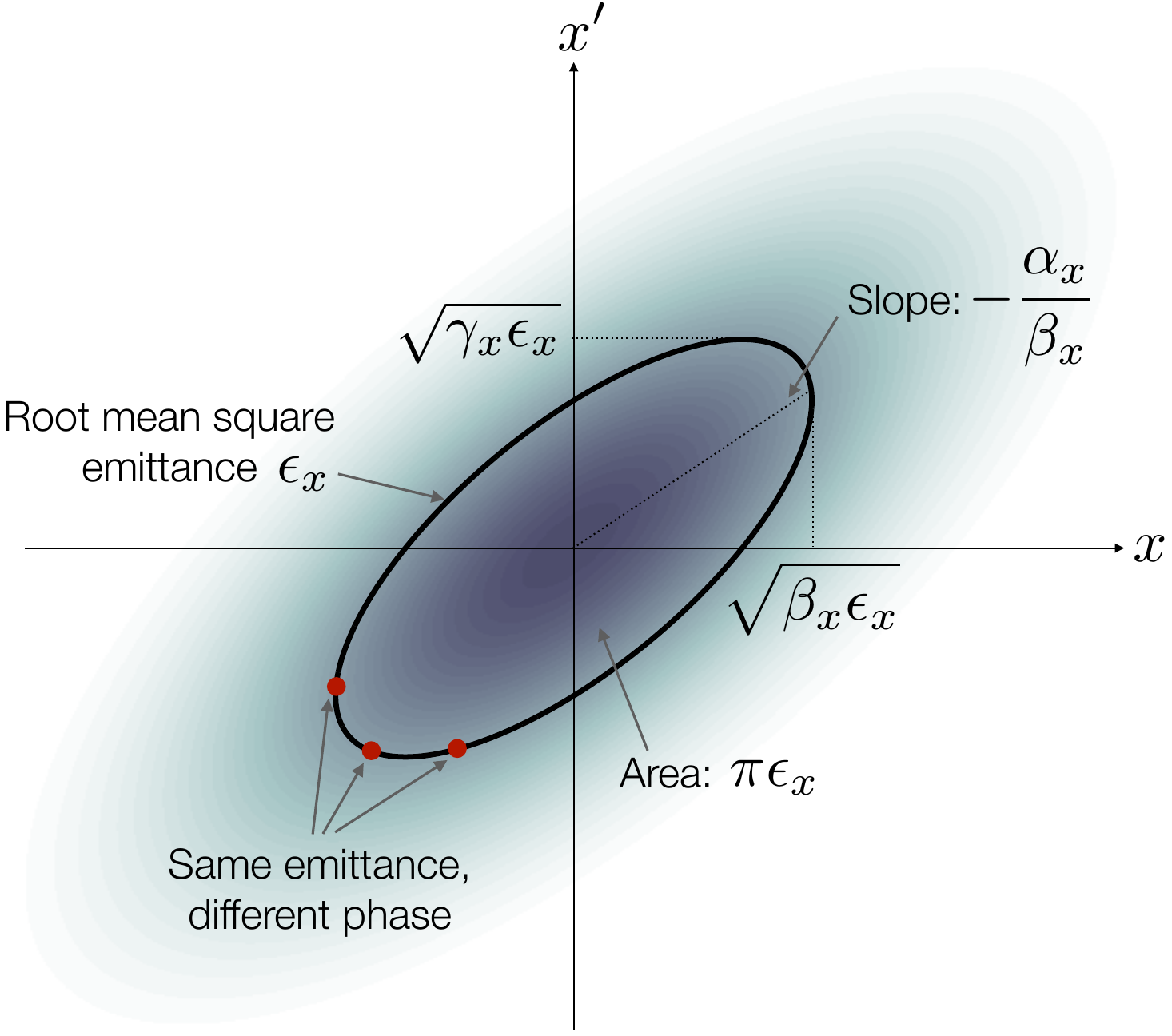}
    \caption{Particle distribution in trace space (non-normalized phase space), with an ellipse defined by particles of the same single-particle emittance and an area proportional to the rms geometric emittance $\epsilon_x$. Twiss parameters $\alpha_x$, $\beta_x$ and $\gamma_x$ are also defined. Source: Ref.~\cite{Lindstrom2019}.}
    \label{fig:trace-space}
\end{figure}

Emittance is defined as the area of a particle distribution in transverse phase space (i.e., particle position, $x$, versus momentum, $p_x$). In simple terms, this is a measure of the beam's ability to be focused to small beam sizes---the lower the emittance, the smaller the beam size throughout a given focusing system. More precisely, the root-mean-square (rms) emittance can be calculated as \cite{Floettmann2003}
\begin{equation}
    \epsilon_n = \frac{\sqrt{\det(\Sigma)}}{m c} = \frac{1}{m c} \sqrt{\langle x^2 \rangle \langle {p_x}^2 \rangle - \langle x p_x \rangle^2},
\end{equation}
where $\Sigma = \mathrm{cov}(x, p_x)$ is the beam's covariance matrix, $\langle \rangle$ denotes the average over all the particles in the distribution, $m$ is the particle mass, and $c$ is the speed of light in vacuum. This \textit{normalized} emittance (as denoted by subscript $n$) is unaffected by pure acceleration, as opposed to the \textit{geometric} emittance (subscript $g$)
\begin{equation}
    \epsilon_g = \sqrt{\langle x^2 \rangle \langle {x'}^2 \rangle - \langle x x'\rangle^2} \approx  \frac{\epsilon_n}{\langle\beta\gamma\rangle},
\end{equation}
which measures the area in so-called \textit{trace space} (particle position, $x$, versus angle, $x'$), as shown in Fig.~\ref{fig:trace-space}, where $\gamma$ and $\beta$ are the relativistic Lorentz factor and the normalized velocity, respectively. The relation between $\epsilon_g$ and $\epsilon_n$ is only exact if the energy spread is negligible. Since the geometric emittance decreases with acceleration, the normalized emittance is the quantity of interest when considering emittance growth and preservation in an accelerator.

Liouville's theorem states that the area of a distribution in phase space, and hence the normalized emittance, can at best stay preserved and otherwise only increase. Such preservation only occurs in the case of linear transformations of the phase space; or in the case of particle beams, whenever linear focusing (or no focusing) is applied. Nonlinear focusing and other influences generally tend to increase the emittance.

In a particle bunch of finite length, two types of emittance can be considered: (1) the emittance of an individual longitudinal slice, known as the \textit{slice} emittance, and (2) the average emittance of all slices together, known as the \textit{projected} emittance. The projected emittance can equal the slice emittance, but will in general be larger when the beam properties or the focusing properties depend on the longitudinal position in the beam.

Finally, it is useful to consider the combined emittance of a distribution of bunches. Averaging the transverse phase space over many shots, we can get an effective emittance known as the \textit{multi-shot} emittance. This is a particularly useful quantity in particle colliders, where the collision rate depends not primarily on the beam size of individual bunches, but the average beam size over millions of shots---if the mean transverse position of the beam jitters enough to only rarely hit the counter-propagating beam, few collisions will occur.

\subsection{Requirements from applications}

Different accelerator applications have drastically different needs. For example, medical applications, sample irradiation~\cite{Faure2016}, or Compton scattering~\cite{Esarey1993} have relatively modest requirements on beam emittance, mostly reachable by today's experiments. On the other hand, free-electron lasers (FELs) and linear colliders require very high beam quality. Because of the potential impact of building smaller and cheaper versions of these large machines, much research is naturally directed toward meeting their beam-quality requirements.

Before discussing the requirements on emittance, it is important to stress that machines such as free-electron lasers and linear colliders need more than just a preserved emittance. In each stage of the accelerator, the beam charge, the energy spread, the bunch length, and potentially even the beam spin polarization~\cite{Vieira2011} also needs to be preserved. Ultimately, what needs to be preserved is the beam density in the full six-dimensional phase space, known as the \textit{6D brightness}
\begin{equation}
    B_{\mathrm{6D}} = \frac{Q}{\sqrt{\det{(\Sigma_{\mathrm{6D}}})}} \approx \frac{Q}{\epsilon_{nx}\epsilon_{ny}\epsilon_{nz}},
\end{equation}
where $Q$ is the bunch charge, $\Sigma_{\mathrm{6D}}$ is the full 6D beam covariance matrix, and $\epsilon_{nz}$ is the emittance in the longitudinal phase space, which is proportional to both the bunch length and the energy spread. The approximation is only exact if there are no cross-plane couplings. Brightness preservation is a much more challenging task than preserving the individual beam qualities, and may therefore impose additional constraints on the design of the accelerator. In this paper, however, we will focus mainly on the preservation of emittance: perhaps the most difficult part, but nevertheless one of many.

In an FEL, the required 6D brightness is of order $10^{16}$~A m$^{-2}$~rad$^{-2}$ per 0.1\% energy bandwidth, or higher. To achieve this, the emittance required is approximately 1~mm-mrad or smaller. Note that this is the requirement on the slice emittance---the projected emittance can be larger in certain cases, for instance in the case of chromatic emittance growth where a flying focus can be formed~\cite{Huang2012}. 

In a linear collider, there is no direct dependence on the 6D brightness; what matters is how small the beam size can be made at the collision point. The collision rate is proportional to the luminosity
\begin{equation}
    \mathcal{L} = H_{\mathrm{D}}\frac{N^2 f \gamma}{4 \pi \sqrt{\beta_x \epsilon_{nx}} \sqrt{\beta_y \epsilon_{ny}}},
\end{equation}
where $N$ is the particle number, $f$ is the repetition rate, $\beta_{x,y}$ is the Twiss beta function at the interaction point, and $H_{\mathrm{D}}$ is a numerical factor. Due to constraints from the final focusing system as well as from beam--beam effects (beamstrahlung)~\cite{Yokoya1992}, the emittance of the beam is usually required to be asymmetric (flat): 1--10~mm-mrad in the horizontal plane, and as low as 0.01~mm-mrad in the vertical plane~\cite{ILC2013,CLIC2013}. This very small vertical emittance is a particularly challenging goal, both for conventional and for advanced accelerators. Symmetric (round) beams may be viable if the bunch length is sufficiently short to suppress beamstrahlung, which would loosen the emittance growth tolerance in the vertical plane---an option compatible with many advanced accelerator concepts.

Since emittance never decreases in a single-pass accelerator, it is meaningful to think of the difference between the initial emittance and the required final emittance in the form of an \textit{emittance budget}. In practice, we cannot avoid all emittance growth, so we need to manage and distribute it carefully. The question will be how to balance the various sources of emittance growth in the accelerator: both inside the stages, outside the stages, and the interplay between the two.

\section{Sources of emittance growth in an accelerating stage}
\label{sec:emittance-growth-in-stage}

In the most general sense, there are two reasons why emittance preservation inside an advanced accelerator is more difficult than in a conventional accelerator. Firstly, the operating wavelength is much smaller: whereas conventional RF accelerators typically have cavities on the 100~mm scale, the accelerating structures in advanced accelerators typically range from 1~mm to 1~{\textmu}m. This is problematic because the accelerating bunch now takes up a larger volume relative to the volume of the accelerating structure, making it more difficult to have uniform longitudinal and transverse wakefields, as well as making it more difficult to precisely align the beam to the center. It also means that the accelerating bunch has a higher charge per surface area of the accelerating structure, which can lead to problems of instability. Secondly, though only in the case of plasma-based accelerators, there can be on-axis ions and electrons, which on the one hand can be beneficial for focusing and stability, but on the other hand can also introduce new sources of emittance growth. 

This section covers the most important sources of emittance growth inside an advanced accelerator stage. We start by considering general issues that apply to all advanced accelerators, including space charge and instabilities due to transverse wakefields, and then continue with issues specific to plasma-based accelerators, including energy-spread-based decoherence effects, nonuniform focusing fields, Coulomb scattering, and radiation-based effects.

\subsection{Space charge}
Space charge is a basic collective effect whereby the particles are repelled by the charge of the other particles in the bunch. In principle, this happens in all accelerators, but the effect is often negligible, especially in plasma accelerators~\cite{Shanks2009}. The exception is accelerators in which beams with both low beam energy and high charge density occur, such as in the initial stages of a dielectric laser accelerator~\cite{England2014}.

How strong is the self-defocusing force from space charge? Consider the radial envelope equation for a charged particle beam~\cite{Reiser1994},
\begin{equation}
    r'' + \frac{\gamma'}{(\beta \gamma)^2} r' - \frac{{\gamma'}^2}{(\beta \gamma)^2} r + k_r r  - \frac{K}{r} - \frac{\epsilon_n^2}{(\beta \gamma)^2 r^3} = 0,
\end{equation}
where $r$ denotes the envelope radius, and $k_r$ is the radial focusing strength. Here, $K$ is the so-called \textit{generalized perveance}, which describes the space-charge force, and is defined as
\begin{equation}
    \label{eq:generalized-perveance}
    K = \frac{2}{(\beta \gamma)^3} \frac{I}{I_A},
\end{equation}
where $I$ is the beam current, and $I_A \approx 17$~kA is the Alfv{\'e}n current. The space-charge term in the envelope equation is negative (i.e., defocusing) and scales inversely with the beam radius---smaller beam sizes are therefore more strongly affected. Equation~\ref{eq:generalized-perveance} also shows that the defocusing force scales with the beam current and that it is strongly suppressed at higher energies. This suppression can be understood by considering the bunch in different reference frames: higher energy results in more relativistic length contraction in the lab frame, which implies a lower charge density and therefore less space charge in the rest frame.

While the defocusing can be compensated by additional focusing (through the $k_r r$ term), in general the space-charge force is not emittance preserving. Only in the case of a radially uniform charge-density distribution (a so-called \textit{K-V distribution} after Kapchinsky and Vladimirsky~\cite{Kapchinsky1959}) would the emittance be preserved; not the case for the more regular Gaussian distribution~\cite{Lagniel1994}. In short, significant space charge should be avoided throughout the accelerator.

\subsection{Transverse wakefields and instabilities}

\begin{figure}[t]
    \centering
    \includegraphics[width=0.55\linewidth]{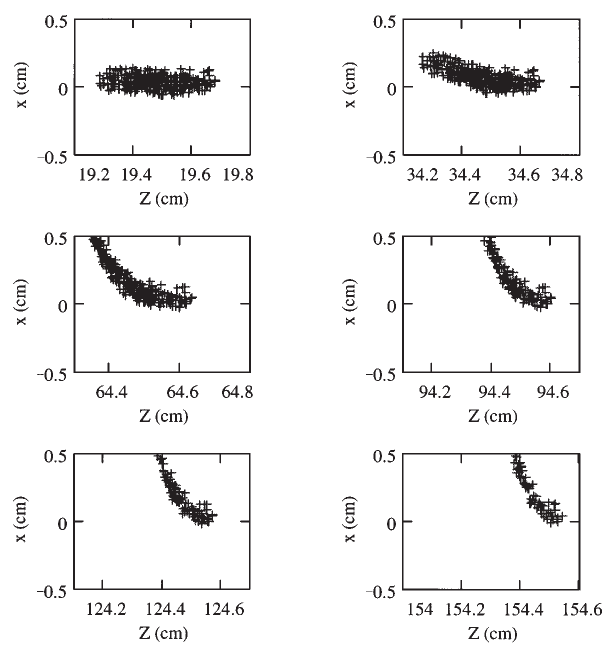}
    \caption{Instability due to transverse wakefields in a dielectric wakefield channel with no external focusing, growing along the accelerator from an initial transverse $x$ offset (upper left) to result in a significant charge loss (lower right), where $Z$ refers to the longitudinal propagation distance. In this simulation, the bunch charge is 40~nC and the energy 150~MeV, and the dielectric tube has a radius of 5~mm (the edge of the simulation box). Source: Ref.~\cite{Gai1997} (reproduced with permission).}
    \label{fig:bbu}
\end{figure}

Another collective effect that is common to all accelerators is the transverse wakefield. The longitudinal accelerating field makes up the fundamental mode of the wakefield in an accelerating structure, but further modes are also possible. The transverse wakefield appears whenever the beam is off axis---it scales with the transverse offset---and is a deflecting force that pulls particles away from the axis. The \textit{short-range wake theorem}~\cite{Fedotov1999,Lebedev2017} links this transverse wakefield per offset $W_x$ to the longitudinal wakefield $W_z$ directly behind a particle (assuming cylindrically symmetric geometry):
\begin{equation}
    \label{eq:short-range-wake-theorem}
    W_x(z) \approx \frac{2}{a^2} \int_0^z W_z(z')dz',
\end{equation}
where $a$ is the characteristic cavity dimension and $z$ is the (co-moving) distance behind the particle. The wakefield is defined as the force per charge, meaning that the deflecting force scales linearly with both the transverse offset and the charge. In principle, the transverse wakefield can be suppressed in the case of perfectly on-axis beams. However, since even an infinitesimal offset leads to further deflection, exponential growth occurs and the beam is lost (see Fig.~\ref{fig:bbu}).

Importantly, Eq.~\ref{eq:short-range-wake-theorem} shows why transverse wakefields are a major focal point for advanced accelerators: the transverse wakefield scales with the inverse square of the cavity dimension $a$ compared to the strength of the accelerating field. This means that accelerators with a small accelerating structure, which goes for all advanced accelerators, have much stronger transverse wakefields and are therefore more prone to instability. The transverse wakefield can be somewhat suppressed in a highly asymmetric (planar) geometry \cite{O'Shea2020}, where the scaling with $a$ in Eq.~\ref{eq:short-range-wake-theorem} does not hold \cite{Baturin2018}---this can be an option in dielectric structures, but will likely be challenging in plasmas.

An example of this effect can be seen in hollow plasma channel accelerators~\cite{Tajima1983}, where a leading driver sets up a strong longitudinal wakefield in a hollow tube of plasma---a concept that has been investigated for its possibility for accelerating positrons at high gradient~\cite{Lee2001}. However, the transverse wakefield that results from a misaligned bunch results in catastrophic beam loss, even in a short accelerator stage~\cite{Schroeder1999,Gessner2016}. This was recently demonstrated in an experiment at SLAC~\cite{Lindstrom2018a}.

In order to avoid such deflection, external or internal focusing is required. At a minimum, this focusing force per offset must equal that of the transverse wakefield: $k_r \geq Q W_x$. Unfortunately, even in the presence of focusing, instability can occur due to a resonance between the different slices of the beam---an effect known as the \textit{beam-breakup instability} \cite{Neil1979,Chao1980,Lau1989}.

\subsubsection{Beam-breakup instability}
\label{sec:beam-breakup-instability}
The beam-breakup instability was discovered early in the history of particle accelerators. Kelliher and Beadle reported already in 1960 that the bunch length of an electron beam would shorten by itself~\cite{Kelliher1960}. This was also observed in 1966 in the SLAC linac, and subsequently explained by Panofsky and Bander as being caused by "radial modes" (transverse wakefields) in the beam pipe, driven by the bunch head and causing a deflection of the bunch tail~\cite{Panofsky1968}.

To understand how the instability occurs, it is instructive to consider the system in its simplest form: a two-particle model. The leading particle has a charge $q_1$ and an offset $x_1$, and is oscillating in a focusing channel (e.g., of external quadrupoles) with a wavenumber $k_{\beta}$:
\begin{equation}
    x_1(s) = x_0 \cos(k_{\beta}s),
\end{equation}
where $x_0$ is the initial offset and $s$ is the longitudinal coordinate along the beamline. This results in a trailing transverse force $F_x = q_1 x_1 W_x$ observed by the second particle, whose offset is $x_2$. This trailing particle, which stays at a fixed distance $\Delta z$ behind the leading particle, is experiencing both the focusing channel and the transverse wakefield, and is therefore described by the equation for an externally driven harmonic oscillator,
\begin{equation}
    \label{eq:two-particle-model}
    \frac{\partial^2 x_2}{\partial s^2} + k_{\beta}^2 x_2 = \frac{q_1 x_0}{E} W_x(\Delta z) \cos(k_{\beta}s),
\end{equation}
where $E$ is the energy of the trailing particle. The solution to Eq.~\ref{eq:two-particle-model} is given by
\begin{equation}
    x_2(s) = x_0 \cos(k_{\beta}s) + \frac{q_1 x_0}{2 k_{\beta} E} W_x(\Delta z) s \sin(k_{\beta}s),
\end{equation}
where the second term has a linearly growing amplitude (scaling with $s$) due to resonance. Since the transverse wakefield initially increases linearly with $\Delta z$, and the wakefields of successive longitudinal slices add together, the tail of the bunch is likely to see the largest amplitude oscillations, leading to beam breakup and a progressive shortening of the bunch.

Suppressing the instability requires breaking the resonance, which means that each longitudinal bunch slice needs to oscillate at a different frequency in the focusing channel. The most straightforward technique is known as \textit{BNS damping} after Balakin, Novokhatsky and Smirnov's suggestion in 1983~\cite{Balakin1983}, which involves imprinting an energy chirp along the bunch---if the tail has a slightly lower energy than the head, it will oscillate at a slightly higher frequency in the focusing channel. Returning to the two-particle model, this implies that the focusing channel has a different strength $k_{\beta}+\Delta k_{\beta}$ for the second particle, such that the differential equation becomes
\begin{equation}
    \frac{\partial^2 x_2}{\partial s^2} + (k_{\beta}+\Delta k_{\beta})^2 x_2 = \frac{q_1 x_0}{E} W_x(\Delta z) \cos(k_{\beta}s),
\end{equation}
which has the solution (when $\Delta k_\beta \neq 0$)
\begin{equation}
    x_2(s) = x_0 \cos\left((k_{\beta}+\Delta k_{\beta})s\right) + \frac{q_1 x_0}{2 k_{\beta} \Delta k_{\beta} E_2} W_x(\Delta z) \left(\cos\left((k_{\beta}+\Delta k_{\beta})s\right) - \cos(k_{\beta}s)\right).
\end{equation}
This oscillation is not resonant, which means that the instability is suppressed.

Beam breakup is one of the main problems in wakefield accelerators using dielectric structures. Applying strong external focusing via quadrupoles as well as BNS damping using a chirp can mitigate the problem, as shown by Gai \textit{et al.}~\cite{Gai1997}, but parameter studies have shown that the effect ultimately limits the gradient of such accelerators to at most a few hundred MV/m~\cite{Li2014}---well short of the desired gradient range of GV/m. Hollow channel plasma accelerators, where a plasma is the dielectric medium, are expected to be similarly limited~\cite{Lindstrom2018a}.

\subsubsection{Hose instability}
The \textit{electron-hose instability} in a plasma accelerator, first introduced by Whittum \textit{et al.}~in 1991~\cite{Whittum1991}, is conceptually very similar to the beam-breakup instability, mainly thought of as affecting the driver. Here, the electron bunch experiences the focusing of the on-axis ion column (i.e., internal focusing), as opposed to conventional accelerators where the focusing is done by quadrupoles (i.e., external focusing). If the bunch has a tilt, longitudinal slices will have a slight transverse offset from the axis, which causes the plasma-accelerator cavity behind that point to also be slightly transversely offset. Just like conventional beam breakup, this causes a resonance that can lead to violent instability of the beam driver and hence the cavity in which the trailing bunch sits, causing significant emittance growth or charge loss. What distinguishes the hose instability from the beam-breakup instability is mainly its scalings, because the channel radius depends on the beam parameters---in particular, Whittum \textit{et al.}~showed that this causes hosing to be independent of beam current. However, this simple conclusion was later challenged when the theoretical framework was improved upon by Huang \textit{et al.}~\cite{Huang2007} and Mehrling \textit{et al.}~\cite{Mehrling2017,Mehrling2018a}, showing that hosing does in fact scale with beam current. They also concluded that the instability was less severe than originally anticipated, though still important to mitigate.

\begin{figure}[t]
    \centering
    \includegraphics[width=0.65\linewidth]{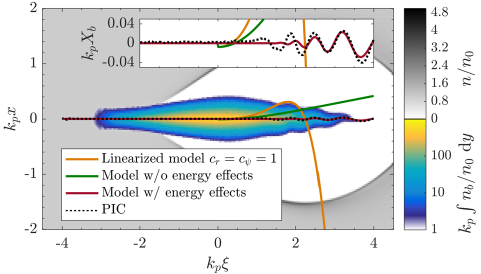}
    \caption{The electron-hose instability, simulated in a particle-in-cell (PIC) code, overlaid with analytical models with a varying level of accuracy. Here, $n_b$, $n$ and $n_0$ are the beam density, local and background plasma density, respectively, $\xi$ is the comoving longitudinal coordinate, $X_b$ is the beam centroid offset, and $c_r$ and $c_{\psi}$ are coefficients accounting for relativistic motion. Source: Ref.~\cite{Mehrling2017} (reproduced with permission).}
    \label{fig:hosing}
\end{figure}

Suppression of the hose instability builds on the tactics from the beam-breakup instability~\cite{Mehrling2019}: break the resonance. Three main directions have been proposed. (1) Firstly, a strong energy chirp can be imposed on the bunch for BNS damping. This is often a side-effect of the strong decelerating fields in the driver, leading to natural suppression of the hose instability in beam drivers~\cite{Mehrling2017}. (2) Secondly, at the head of a beam driver, there is a natural ramp-up of the focusing gradient as the plasma electrons are repelled off axis---especially when the plasma is beam-ionized~\cite{Deng2006}. The same method can also be used in a pre-ionized plasma by employing large-radius beam drivers~\cite{MartinezDeLaOssa2018}, where the beam radius is similar to that of the plasma-wake cavity. Here, there is also a radial focusing gradient that causes further suppression of the resonance. (3) Thirdly, and relatedly, if the driver has a high charge density, it can pull ions onto the axis (known as \textit{ion motion}) such that the on-axis focusing is locally nonlinear~\cite{Mehrling2018b}---avoiding the resonance by avoiding harmonic oscillations. This nonlinear focusing, however, may itself cause emittance growth; a topic that is discussed in Section~\ref{sec:ion-motion}. In all of these solutions, the betatron frequency depends on the comoving coordinate. In (1), this dependence is imprinted on the particle energy; in (2) and (3) it is imprinted on the focusing force.

Finally, the hose instability is not limited to particle beam drivers. The laser pulse in a laser-driven wakefield accelerator can also be subject to hose instability, although this is usually less significant in practice for short laser pulses~\cite{Sprangle1994}. When the accelerating bunch experiences the hose instability, this is often just referred to as a beam-breakup instability (Section~\ref{sec:beam-breakup-instability}), but can be mitigated using many of the same principles as for the beam driver~\cite{Lehe2017}.

\subsection{Energy spread and decoherence}
For most applications, bunches with zero energy spread would be ideal. However, this is not practically feasible in a high-gradient wakefield accelerator due to the high-frequency oscillation of the accelerating field---only by carefully shaping the current profile for beam loading~\cite{Tzoufras2008,Lindstrom2021a,Kirchen2021} or by periodically modulating the density~\cite{Brinkmann2017} can the wakefield be approximately flattened. On the flip side, in order to suppress instabilities we can benefit from a non-zero energy spread, as detailed in the previous section.

In conventional accelerators, and some advanced accelerators, the presence of an energy spread does not significantly impact the transverse phase space and the emittance. However, in plasma-based accelerators with strong on-axis focusing, it does: particles of different energy will oscillate with different frequency through the focusing channel, leading to \textit{decoherence} (or \textit{phase mixing}) of the energy slices of the beam. This causes a projected emittance growth unless the beam is both perfectly \textit{matched} and \textit{aligned}, as covered by the sections below.

\subsubsection{Mismatching}
\label{sec:mismatching}
Matching is a generic term in beam dynamics that refers to achieving a specified set of Twiss parameters (i.e., alpha function $\alpha_x$ and beta function $\beta_x$). In reality, we will never reach complete matching, so it can be useful to quantify the level of mismatch. This is often done with the \textit{mismatch parameter}~\cite{Sands1991}
\begin{equation}
    \mathcal{M} = \frac{1}{2}\left( \tilde{\beta} + \tilde{\gamma} + \sqrt{(\tilde{\beta} + \tilde{\gamma})^2 - 4} \right),
\end{equation}
where $\tilde{\beta} = \beta_x/\beta_m$, $\tilde{\alpha} = \alpha_x - \alpha_m\tilde{\beta}$, and $\tilde{\gamma} = (1+\tilde{\alpha}^2)/\tilde{\beta}$ are the normalized errors of the Twiss parameters, and the index "$m$" indicates matched parameters. 

In the ion column of a plasma accelerator, the matched beta function is defined as that which stays constant throughout the channel. We can find this by setting all derivatives to zero in Hill's equation, which is given by
\begin{equation}
    \frac{1}{2}\beta(s)\beta''(s) - \frac{1}{4}\beta'(s)^2 + K(s) \beta(s)^2 = 1,
\end{equation}
resulting in $\beta_m = 1/\sqrt{K}$, as well as $\alpha_m = -\beta'(s)/2 = 0$. For a plasma, the focusing constant $K = 2E\epsilon_0/n e^2$, where $n$ is the exposed charge number density in the ion column, and $\epsilon_0$ and $e$ are the vacuum permittivity and elementary charge, respectively. For a plasma accelerator operating in the blowout regime, this can be recast as 
\begin{equation}
    \beta_m = \frac{\sqrt{2\gamma}}{k_p},
\end{equation}
where $k_p = \sqrt{n e^2/ \epsilon_0 m_e c^2}$ is the plasma wavenumber, in which $m_e$ and $c$ are the electron mass and speed of light in vacuum.

\begin{figure}[b]
    \centering
    \includegraphics[width=0.6\linewidth]{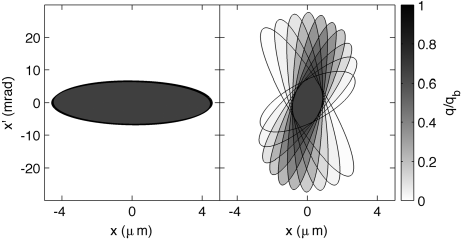}
    \caption{After initially being coherent, but mismatched to the ion column (left panel), different energy slices gradually decohere in phase space (right panel). The shading indicates the ratio of the slice charge $q$ to the total bunch charge $q_b$. Source: Ref.~\cite{Mehrling2012} (CC BY 3.0).}
    \label{fig:mismatching}
\end{figure}

In the presence of an energy spread, the energy slices of the bunch will rotate in phase space at different rates. If all the energy slices are matched, this is inconsequential since the distribution rotates in phase space without changing shape. However, if the bunch is mismatched, the energy slices will gradually decohere, as illustrated in Fig.~\ref{fig:mismatching}.

Eventually, the energy slices will fully decohere: at that point the beta function returns to the matched beta function, but the emittance has grown. This saturated emittance growth can be expressed purely in terms of the mismatch parameter:
\begin{equation}
    \frac{\epsilon_{\mathrm{sat}}}{\epsilon_{\mathrm{init}}} = \frac{1}{2}\left(\mathcal{M} + \frac{1}{\mathcal{M}} \right),
\end{equation}
where $\epsilon_{\mathrm{init}}$ is the initial emittance. How long does it take for the energy slices to fully decohere? The \textit{complete decoherence length}~\cite{Mehrling2012} is given by
\begin{equation}
    \label{eq:decoherence-length}
    L_{\mathrm{dc}} = \frac{\lambda_{\beta}}{\sigma_{\delta}}
= \frac{2 \pi \beta_m}{\sigma_{\delta}},
\end{equation}
where $\sigma_{\delta}$ is the rms relative energy spread, and $\lambda_\beta$ is the betatron wavelength. In other words, complete decoherence occurs after $1/\sigma_{\delta}$ betatron oscillations. As an example, consider a plasma column of density $10^{17}$~cm$^{-3}$ and an incoming beam of energy 1~GeV, 3\% energy spread, 5~mm beta function, and alpha function of $-1$; the matched beta function is 1.05~mm, and therefore the mismatch parameter would be 4.98, resulting in a saturated emittance growth of 259\% after a decoherence length of 0.22~m (33 betatron oscillations). Another, more application-relevant example was simulated by Thomas and Seipt~\cite{Thomas2021}, showing that acceleration to 1~TeV across 85 stages would require an initial energy spread below 0.1\% and a growth below 0.001\% of the energy gain per stage in order to avoid significant emittance growth.

Reaching the small beta functions required for matching can be challenging in a real accelerator. A solution to this problem is to locally increase the matched beta function at the entry and exit of the plasma accelerator by way of a longitudinal \textit{plasma-density ramp}~\cite{Marsh2005,Dornmair2015,Xu2016,Ariniello2019}. This allows the beta function to gradually decrease to and increase from the required value, helping to preserve the emittance when coupling in and out of stages---a topic that is covered in Section~\ref{sec:emittance-growth-between-stages}.

\subsubsection{Transverse misalignment}

\begin{figure}[t]
    \centering
    \includegraphics[width=0.6\linewidth]{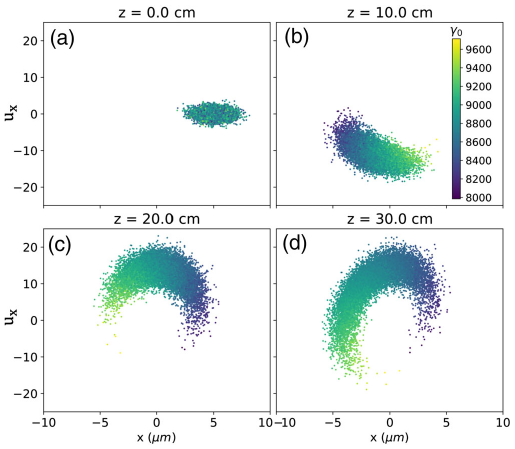}
    \caption{Transverse misalignment of a beam with a finite energy spread, evolving in phase space as it traverses a laser-plasma accelerator from the start (a) to 30 cm downstream (d). As the different energy slices rotate at different rates, they decohere and cover a larger area in phase space---an emittance growth. Here, $u_x = p_x/(m_e c)$ is the transverse normalized momentum, $\gamma_0$ is the longitudinal Lorentz factor of the particles, and $z$ is the longitudinal position. Source: Ref.~\cite{Thevenet2019} (CC BY 3.0).}
    \label{fig:misalignment}
\end{figure}

An issue related to mismatching is that of misalignment: in many ways it is the same problem, one azimuthal mode up. If the bunch is off-axis in the focusing channel, different energy slices will rotate at different rates, causing them to decohere and spread out in phase space (see Fig.~\ref{fig:misalignment}). The resulting particle distribution is a ring in phase space, convolved with the original particle distribution, with an emittance increase given by
\begin{equation}
    \label{eq:misalignment}
    \Delta\epsilon_{\mathrm{misalign}} \approx \frac{1}{2}\left(\frac{\Delta x^2}{\beta_m} + \beta_m\Delta{x'}^2 \right),
\end{equation}
where $\Delta x$ and $\Delta x'$ are the positional and angular offsets from the axis. This emittance growth is only reached after full decoherence, which occurs over a length $L_{\mathrm{dc}}$ (see Eq.~\ref{eq:decoherence-length}). 

In an accelerator with multiple stages, misalignment is applied repeatedly. Initially, where the beam energy is low, the decoherence length is short (due to a small matched beta function): this causes the beam to fully decohere in every stage. At some point, the decoherence length becomes longer than the stages themselves, leading to a complicated "random walk" in the phase space. Both these effects scale as $\sqrt{N_{\mathrm{stages}}}$, where $N_{\mathrm{stages}}$ is the number of stages. Several studies have simulated such multistage accelerators~\cite{Assmann1998,Cheshkov2000,Schulte2016,Lindstrom2016a,Thevenet2019}, concluding that for collider-relevant parameters the required alignment tolerances are of order 10~nm and 1~{\textmu}rad rms. Reaching these extremely tight tolerances will be a major challenge in future advanced accelerator research.

\subsection{Nonlinear focusing fields}

Liouville's theorem dictates that nonlinear transformations of a phase space will cause the emittance to increase. Nonlinear focusing is not very common in conventional accelerators beyond sextupole and octupole magnets, but can easily occur in a plasma accelerator. Such nonlinear focusing can be due to the shape of the wake (Section~\ref{sec:wake-nonlinear}) or due to the response of the plasma to the accelerated beam (Secs.~\ref{sec:ion-motion} and \ref{sec:transverse-beam-loading}).

\subsubsection{Wakes with nonlinear focusing fields}
\label{sec:wake-nonlinear}

In the linear and quasilinear regimes of plasma acceleration, the focusing force is the product of a linear term with the driver's transverse envelope, causing the focusing force to be linear only near the propagation axis. In practice, this is rarely a limitation as the trailing beam is usually much thinner than the driver. In the blowout regime, the focusing force on a trailing electron beam is linear provided the trailing beam remains in the ion cavity, regardless of the beam charge. The acceleration of a positron beam in a nonlinear wake cannot exploit this property: in a number of schemes proposed today, the region with accelerating fields for positrons demonstrates a strongly nonlinear focusing force~\cite{Corde2015,Diederichs2019}. While it is impossible for emittance to be fully preserved in these cases, the emittance grows to reach an equilibrium, often with a complex non-Gaussian phase space~\cite{Lotov2017}. This means that whenever nonlinear focusing is either useful in itself (e.g., for suppressing the beam-breakup instability) or it is a necessary side effect of a scheme (e.g., for positron acceleration), it may be worth "paying the price" of some emittance growth in order to reach the overarching goals.

\subsubsection{Ion motion}
\label{sec:ion-motion}

\begin{figure}[b]
    \centering
    \includegraphics[width=0.6\linewidth]{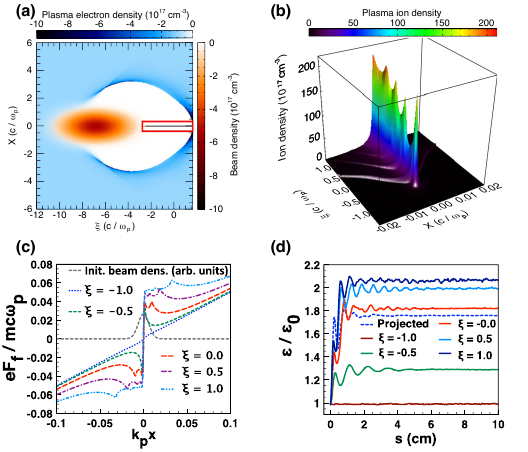}
    \caption{Ion motion in a beam-driven plasma accelerator, resulting in an on-axis ion-density spike (a--b) behind the driver. Nonlinear focusing around the axis (c) causes a longitudinally varying emittance growth (d). Here, $F_f$ denotes the transverse deflecting force, $\omega_p$ is the angular plasma frequency, $\xi$ is the comoving longitudinal coordinate, $\epsilon/\epsilon_0$ is the final-to-initial emittance ratio, and $s$ is the longitudinal position along the plasma accelerator. Source: Ref.~\cite{An2017} (reproduced with permission).}
    \label{fig:ion-motion}
\end{figure}

\begin{figure}[t]
    \centering
    \includegraphics[width=0.7\linewidth]{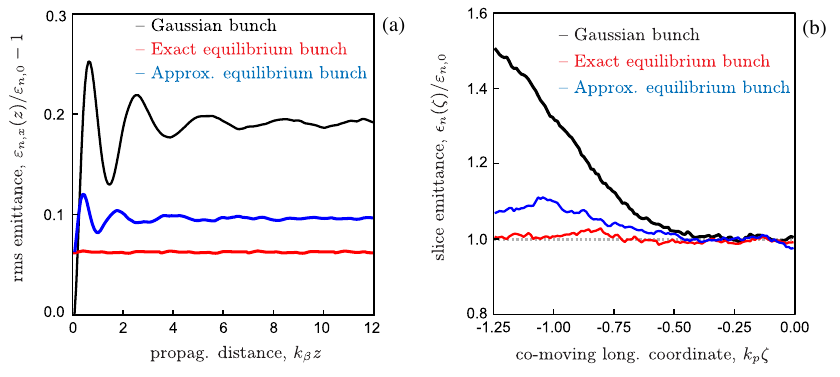}
    \caption{Emittance preservation in the presence of ion motion, achieved by initializing the beam with its equilibrium charge-density profile. In (a), projected emittance growth is plotted versus propagation distance $z$ shown in units of betatron skin depths ($1/k_{\beta}$), and in (b), the slice emittance growth is for different slices at comoving longitudinal position $\xi$ in units of the plasma skin depths ($1/k_p$). Source: Ref.~\cite{Benedetti2017} (CC BY 4.0).}
    \label{fig:ion-motion-emittance-preservation}
\end{figure}

Ions are often assumed to be stationary in a plasma-wakefield accelerator. However, if the electron beam is sufficiently dense, the radial impulse imparted by the beam will move the ions noticeably within one plasma-electron oscillation. This ion motion was first discussed by Rosenzweig \textit{et al.}~\cite{Rosenzweig2005}, who calculated that within a bunch of length $\sigma_z$, the ions would oscillate through a phase advance of
\begin{equation}
    \label{eq:ion-motion-phase-advance}
    \Delta\phi \approx \sqrt{\frac{2 \pi r_a Z \sigma_z N}{A\epsilon_n}}\left(r_e n \gamma \right)^{1/4},
\end{equation}
assuming a matched beam with $N$ electrons, where $\epsilon_n$ is the normalized emittance, $\gamma$ is the beam's Lorentz factor, $n$ is the plasma density, $Z$ is the ion charge state, $A$ is the atomic mass (in amu), and $r_e$ and $r_a = 1.55\times10^{-18}$~m are the classical electron and singly charged ion radii, respectively. If the phase advance $\Delta\phi$ approaches $\pi/2$, a spike will start to form on axis, as illustrated in Fig.~\ref{fig:ion-motion}\,(a--b). From this, it was concluded that the so-called \textit{afterburner} concept for linear colliders~\cite{Lee2002,Raubenheimer2004} would not work due to catastrophic emittance growth from ion motion. Later studies have found, however, that this need not be the case. A Gaussian bunch will increase its emittance until it reaches a new equilibrium~\cite{An2017}. Better yet, if the beam starts out in the equilibrium state, where each longitudinal bunch slice is tailored to the nonuniform ion focusing, its emittance can remain preserved throughout the length of the stage~\cite{Benedetti2017} (see Fig.~\ref{fig:ion-motion-emittance-preservation}). While this will certainly be complex to achieve experimentally, promising solutions have been proposed~\cite{Benedetti2021}.

Ion motion has a nontrivial interplay with other sources of emittance growth. On the one hand, ion motion can help mitigate hosing and beam-breakup instability. On the other hand, limiting the extent of ion motion can only practically be done by increasing the atomic mass of the ions (see Eq.~\ref{eq:ion-motion-phase-advance}), which may also cause emittance growth due to Coulomb scattering---a topic covered in Section~\ref{sec:scattering} below. Moreover, non-Gaussian equilibrium phase spaces are not easily transported to a downstream stage or application---this would require advanced optics involving achromatic point-to-point imaging between the stages. 

Clearly, to deliver low-emittance electron beams from plasma accelerators, it will be crucial to strike just the right balance between too little and too much ion motion.

\subsubsection{Transverse beam loading}
\label{sec:transverse-beam-loading}

\begin{figure}[b]
    \centering
    \includegraphics[width=0.95\linewidth]{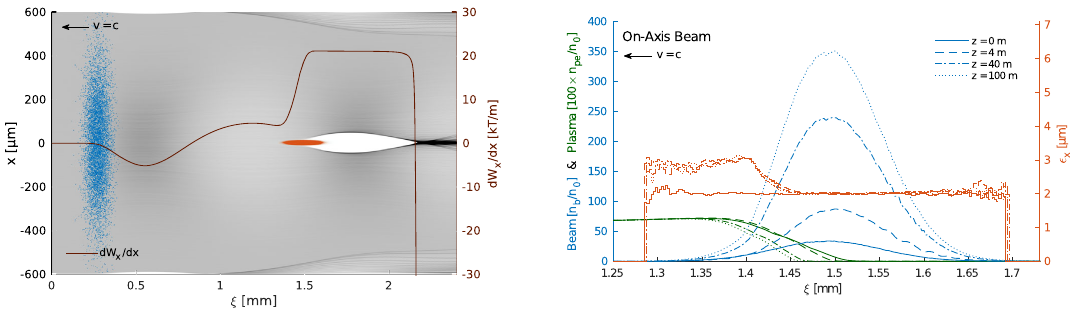}
    \caption{Transverse beam loading in the quasilinear regime, illustrated by a PIC simulation of an electron bunch loading the wakefield driven by a proton driver (left panel). Here, the electron bunch drives its own (strongly nonlinear) wake which loads the wakefield both longitudinally (i.e., flattening the wakefield) as well as transversely (i.e., altering the focusing gradient within the bunch). This causes the head of the bunch to increase its emittance, while the tail is matched and therefore preserves its emittance (right panel). Source: Ref.~\cite{BerglydOlsen2018} (CC BY 4.0).}
    \label{fig:transverse-beam-loading-linear}
\end{figure}

In the linear or quasilinear plasma-wakefield regime, an accelerating bunch that extracts significant energy will not only drive its own longitudinal wake, but also its own transverse wake: this process is called \textit{transverse beam loading}. For instance, an electron bunch will expel plasma electrons from the axis, providing a gradual increase in focusing strength within the bunch, as well as nonlinear focusing fields throughout~\cite{Katsouleas1987}. While this nonlinearity can be useful in damping hosing and beam breakup~\cite{Lehe2017}, it will generally lead to emittance growth. Figure~\ref{fig:transverse-beam-loading-linear} illustrates transverse beam loading in a proton-driven plasma accelerator, where the accelerating electron bunch in a linear wake sees a moderate amount of emittance growth at its head~\cite{BerglydOlsen2018}.

In the context of beam loading, the nonlinear blowout regime is a very special case: if the trailing electron bunch is placed at the back of the bubble, it will beam load the longitudinal wakefield without causing transverse beam loading. The reason is that the trailing bunch can change the trajectory of the in-flowing plasma electrons, but this will not affect the focusing since none of these plasma electrons are on axis. This miraculous fact allows simultaneous high efficiency and low energy spread (from longitudinal beam loading) as well as emittance preservation (from linear ion-only focusing) for electrons accelerated in the blowout regime.

\begin{figure}[b]
    \centering
    \includegraphics[width=0.5\linewidth]{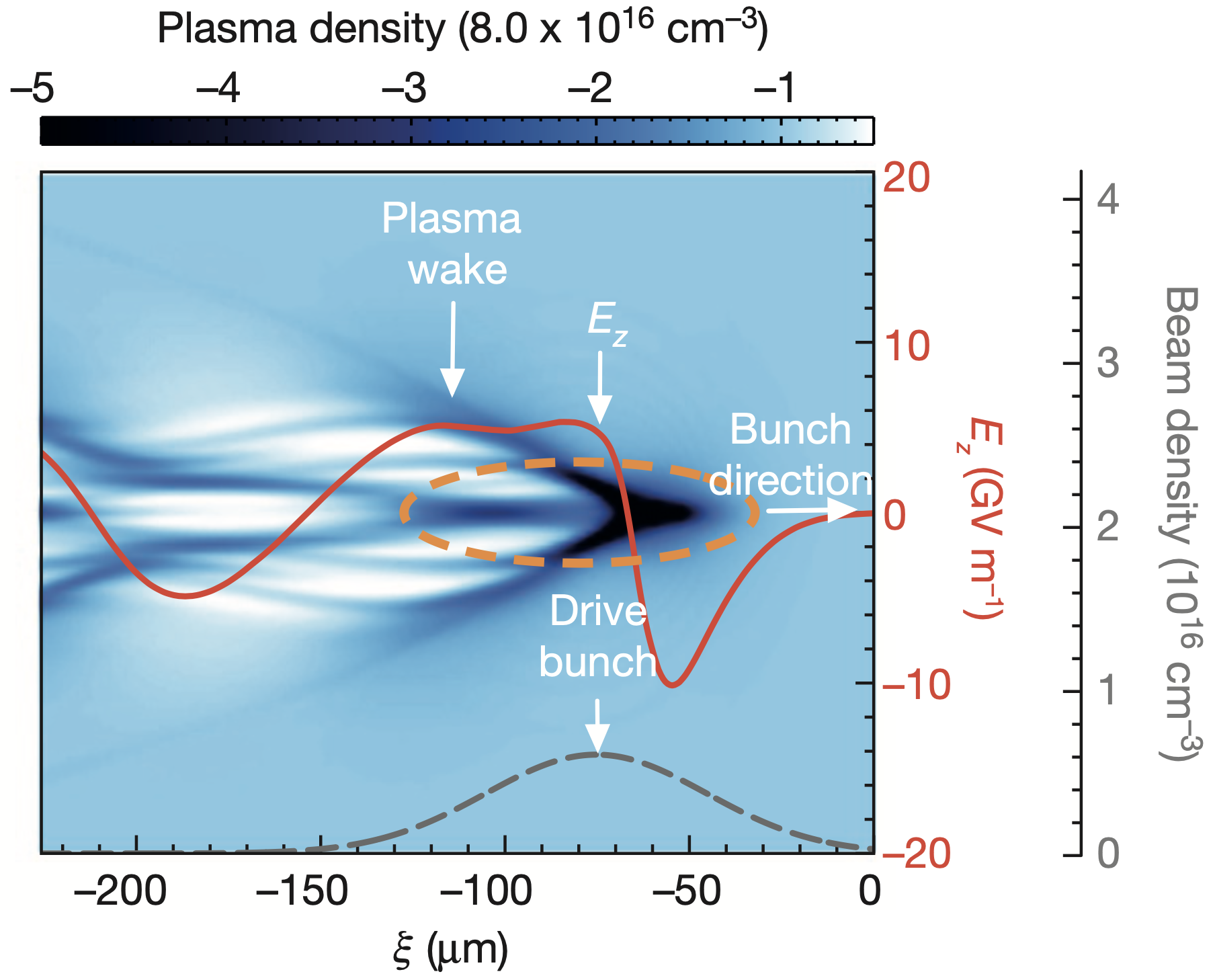}
    \caption{Transverse beam loading in the self-loaded positron acceleration scheme, where plasma electrons are focused onto the axis by the positrons of the bunch. The longitudinal wakefield $E_z$ (red line) is beam loaded by the back half of the positron bunch. The positive charge of the beam also keeps the on-axis electrons from escaping radially, thereby changing the observed focusing field. Source: Ref.~\cite{Corde2015} (reproduced with permission).}
    \label{fig:transverse-beam-loading-positrons}
\end{figure}

Unfortunately, what is a miracle for electron acceleration turns into a curse for positron acceleration. On-axis focusing will be required, otherwise positron bunches will quickly defocus or deflect due to transverse wakefields~\cite{Schroeder1999,Lindstrom2018a}. For this, there must be a negative charge density on axis: a surplus of plasma electrons compared to ions. Simultaneously, to achieve a positive longitudinal wakefield for acceleration, plasma electrons must flow radially outwards. Moreover, high efficiency requires longitudinal beam loading, which means that the trajectory of these electrons will change within the bunch. As a result, efficient positron acceleration in a plasma cannot avoid transverse beam loading. This can be seen in the nonlinear self-loaded regime~\cite{Corde2015} (Fig.~\ref{fig:transverse-beam-loading-positrons}), where a filament of "sucked-in" electrons forms and stays on axis within the strongly beam loading positron bunch. 

To date, no plasma-acceleration scheme for positrons have been shown to provide simultaneous high gradient, high efficiency, low energy spread and emittance preservation. However, promising schemes do exist, including electron-driven wakes in finite-width plasma channels~\cite{Diederichs2019,Diederichs2020}, where a thin on-axis electron-density spike focuses positron beams with minimal emittance growth. This scheme can reach a few percent energy-extraction efficiency, well short of the 50-90\% for blowout regime for electrons, but still reasonable.

\subsection{Coulomb scattering}
\label{sec:scattering}
Conventional accelerators typically operate in an ultra-high vacuum environment---scattering of beam particles off gas atoms occur only rarely. This also goes for most dielectric-based advanced accelerators. In plasma-based accelerators, however, on-axis gas ions and atoms are abundant, and even required for focusing. Coulomb collisions between beam particles and atoms or ions result in either an elastic small-angle deflection of the particle, or (less often) an inelastic collision with energy loss. Assuming mainly elastic scattering, and averaging over many such collisions longitudinally, so-called \textit{multiple Coulomb scattering}, results in an increase in the beam's divergence at a rate~\cite{Montague1984,Kirby2007,Schroeder2013}
\begin{equation}
    \frac{d \langle \theta^2 \rangle}{ds} = \frac{k_p^2 r_e}{\gamma^2}\left[ Z_i^2 \ln\left(\frac{\lambda}{R_a}\right) + 1.78 Z(Z+1) \ln\left(\frac{287}{\sqrt{Z}}\right)\right],
\end{equation}
where $Z$ is the atomic number, $Z_i$ is the ion charge state, $R_a \approx 10^{-10}$~m is the atomic radius, and $\lambda$ is the Debye length (for a neutral plasma) or the plasma wavelength (for an ion channel). Here, the first term is from ion scattering and the second term from neutral atom scattering. For singly gases heavier than hydrogen or helium, the ion-scattering term is usually negligible (assuming singly ionized plasmas). The normalized emittance growth rate can be estimated by 
\begin{equation}
    \frac{d\epsilon_n}{ds} = \frac{\beta_x}{2} \frac{d \langle \theta^2 \rangle}{ds} \gamma,
\end{equation}
where $\beta_x$ is the Twiss beta function. This effect has been included in PIC codes such as WarpX~\cite{Vay2018}, showing emittance growths in line with the analytical model~\cite{Zhao2020}.

Suppressing emittance growth from scattering is most easily done by decreasing the ion mass, by for instance using hydrogen. However, this simultaneously promotes ion motion (Section~\ref{sec:ion-motion}), which also needs to be taken into consideration. Perhaps surprisingly, decreasing the plasma density ($k_p^2 \propto n$) is ineffective, since that results in a lower gradient ($E_z \propto k_p$), hence a longer accelerator, as well as a larger matched beta function ($\beta_m\propto1/k_p$): overall, this cancels out and the integrated emittance growth remains unchanged. 

For typical beam and plasma parameters, including those relevant to linear colliders, the emittance growth due to scattering will be negligible inside the stages, mainly due to the very small beta function~\cite{Kirby2007,Schroeder2010}. However, when considering multiple stages with potentially long plasma-density ramps (Section~\ref{sec:mismatching}) or plasma-lens-based staging optics (Section~\ref{sec:plasma-lenses}), in which the beta function will be significantly larger, scattering may become a problem.

\subsection{Radiative cooling}

In advanced accelerators with strong on-axis focusing, particularly plasma accelerators, synchrotron radiation will be emitted by particles that oscillate off axis~\cite{Esarey2001}. The momentum lost through the emitted photon will be both in the longitudinal and transverse planes. However, the accelerating field is (ideally) purely longitudinal, which implies a damping of the transverse emittance~\cite{Michel2006}. This is how traditional damping rings~\cite{Emma2001} operate---large circular accelerators that produce the lowest emittances available. 

\begin{figure}[t]
    \centering
    \includegraphics[width=0.5\linewidth]{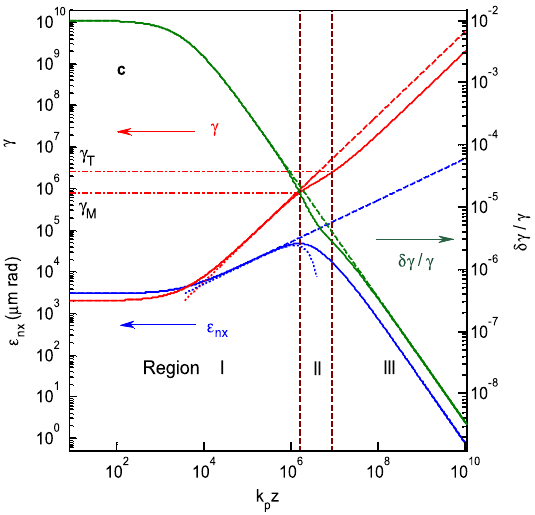}
    \caption{Radiation cooling in a plasma accelerator indicating that, as the bunch is accelerated across a distance $k_p z$, the beam's Lorentz factor $\gamma$ increases (red line), the relative energy spread $\delta\gamma/\gamma$ decreases (green line) and the normalized emittance, after initially increasing, will ultimate be damped by radiation emitted from betatron oscillations (blue line). This calculation assumes ideal conditions such as no transverse misalignment or staging. Source: Ref.~\cite{Deng2012} (CC BY 3.0).}
    \label{fig:radiative-cooling}
\end{figure}

It is currently unclear whether this technique can truly be used for emittance damping in a linear plasma accelerator, due to the limited length (and time) available for damping as well as the complex interplay with the longitudinal phase space, although some authors claim that it could enable low-emittance, low-energy-spread beams well beyond TeV energies~\cite{Deng2012} (see Fig.~\ref{fig:radiative-cooling}).

\section{Sources of emittance growth between stages}
\label{sec:emittance-growth-between-stages}
Connecting multiple accelerator modules may be required for some high-energy applications, such as linear colliders and perhaps free-electron lasers. While this is trivial in conventional RF accelerators---just install modules one after another---this is decidedly not so in many advanced accelerators, particularly plasma-based accelerators~\cite{Lindstrom2021b}. Avoiding staging may be possible, with schemes such as plasma wakes driven by very-high-energy proton bunches~\cite{Caldwell2009,AWAKE2018} or flying-focus laser-plasma acceleration~\cite{Froula2018,Debus2019,Palastro2020}. However, these alternatives are currently severely limited in either energy efficiency or repetition rate. Assuming staging is required, the strong on-axis focusing for transport and suppressing instabilities leads to high beam divergence outside the stage. When combined with large energy spreads or complex non-Gaussian transverse phase spaces, this strong divergence makes it difficult to capture and refocus beams between stages without degrading the beam quality. Emittance growth can then be caused directly by elements used for in- and out-coupling of the driver (Section~\ref{sec:driver-in-and-out-coupling}), by the fact that different energy slices are not focused with the same focal length, an effect known as \textit{chromaticity} (Section~\ref{sec:chromaticity}), or by advanced optical elements like plasma lenses introduced to shrink the distance between the stages (Section~\ref{sec:plasma-lenses}). In the end, the major challenge with staging is to find solutions that preserve emittance and other beam qualities while simultaneously being compact---that is, keeping the \textit{average} acceleration gradient high across the full length of the accelerator.

\subsection{Driver in- and out-coupling}
\label{sec:driver-in-and-out-coupling}
In any wakefield accelerator, the driver, be it a laser pulse or a particle beam, has to be both in- and out-coupled. When operating multiple stages, the trailing bunch must be separated from the depleted driver, which is dumped, and subsequently re-merged with the next driver. Two methods are used: (1) dipole chicanes and (2) mirrors, where the former can be used for all drivers and the latter works only for lasers.

\subsubsection{Magnetic dipole chicanes}

\begin{figure}[b]
    \centering
    \includegraphics[width=0.6\linewidth]{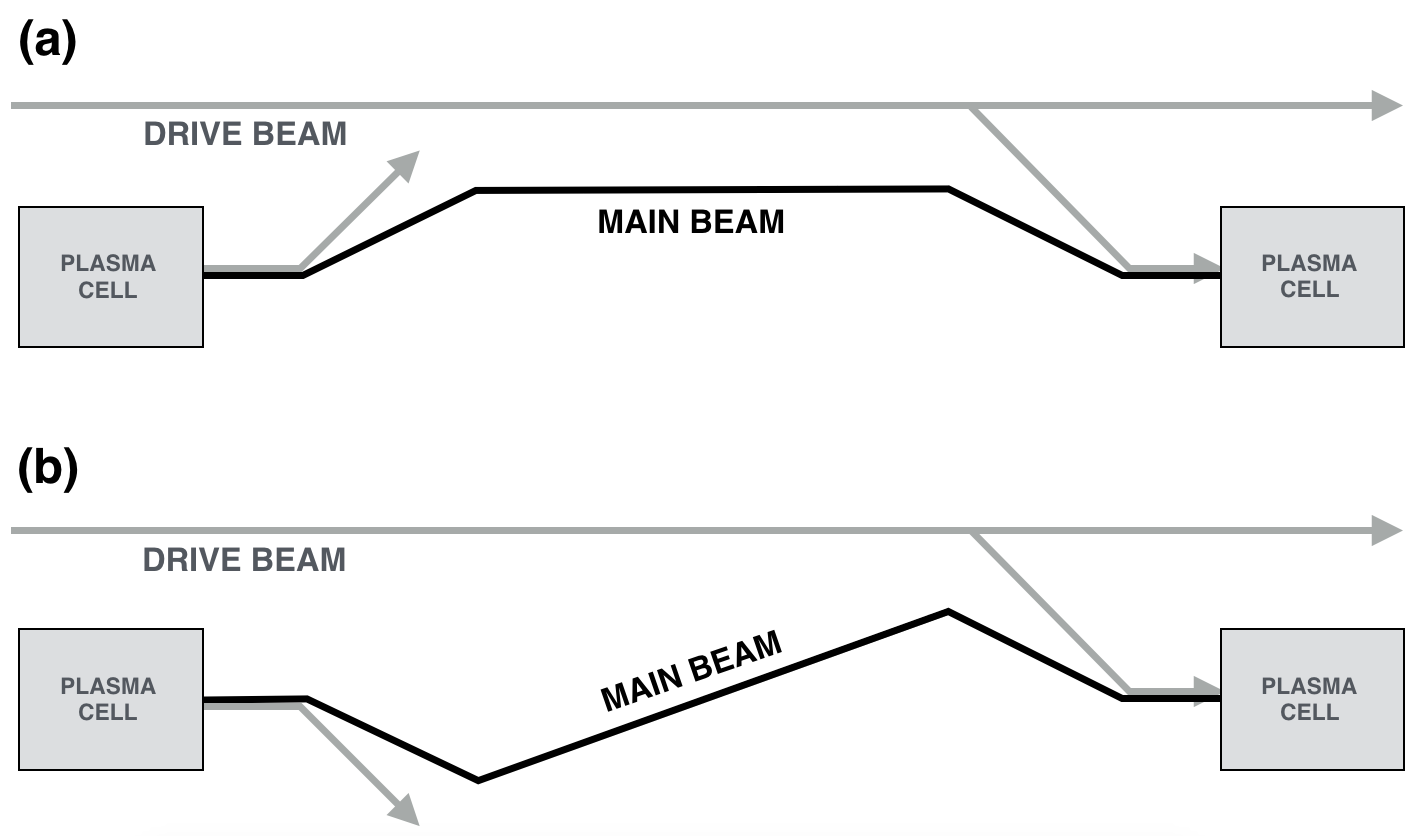}
    \caption{In- and out-coupling of beam drivers using a dipole chicane, showing both a regular chicane (a) and an S-chicane layout (b). A similar scheme could also be used for laser drivers. Here, the bending of the main (trailing) beam will introduce dispersion and coherent synchrotron radiation, which may lead to emittance growth. Source: Ref.~\cite{Lindstrom2016c} (reproduced with permission).}
    \label{fig:dipole-chicanes}
\end{figure}

Dipoles separate beams by charge and by energy: laser drivers will continue straight while the trailing bunch will be deflected; positron and proton beams will be deflected in the opposite direction to an electron bunch; an electron driver can be separated from a trailing electron bunch if their energies do not overlap. The transverse offset resulting from a dipole of length $L$ and magnetic field $B$ is
\begin{equation}
    \Delta x_{\mathrm{dipole}} = \left( \frac{q}{E} - \frac{q_{\mathrm{driver}}}{E_{\mathrm{driver}}} \right) B c L^2,
\end{equation}
where $q_{\mathrm{driver}}$ and $q$ are the elementary charges of the particle species and $E_{\mathrm{driver}}$ and $E$ the energies of the driver and trailing bunches, respectively. Unfortunately, the chicane will also introduce dispersion in any beam with a finite energy spread. Similar to misalignment (Eq.~\ref{eq:misalignment}), this leads to a normalized emittance growth (to lowest order in energy spread) of
\begin{equation}
    \label{eq:dispersion}
    \Delta \epsilon_n \approx \frac{\gamma}{2} \left( \frac{D_x^2}{\beta_m} + \beta_m D_{x'}^2 \right) \sigma_{\delta}^2,
\end{equation}
where $D_x$ and $D_{x'}$ are the first-order positional and angular dispersions, $\beta_m$ is the matched beta function, $\sigma_{\delta}$ is the relative energy spread, and $\gamma$ is the relativistic Lorentz factor. Higher-order dispersion may also need to be considered. As a simple example, consider a 10~GeV beam of 1\% energy spread in a chicane between plasma accelerators of density $10^{17}$~cm$^{-3}$, in which there is a leaking positional dispersion of 10~mm: the emittance would grow by 294~mm-mrad. Moreover, any offset in energy would result in a misalignment, which causes yet more emittance growth. Clearly, dispersion must be cancelled to a very high degree. Plasma density ramps may be used to mitigate the problem, but Eq.~\ref{eq:dispersion} indicates that this merely shifts emittance growth between the $D_x$ and $D_{x'}$ terms. 

Coherent synchrotron radiation~\cite{Nodvik1954,Nakazato1989}, a result of short bunches passing through strong magnetic fields, may introduce further dispersion and emittance growth \cite{Kim2021}.

\subsubsection{Plasma mirrors}

\begin{figure}[b]
    \centering
    \includegraphics[width=0.9\linewidth]{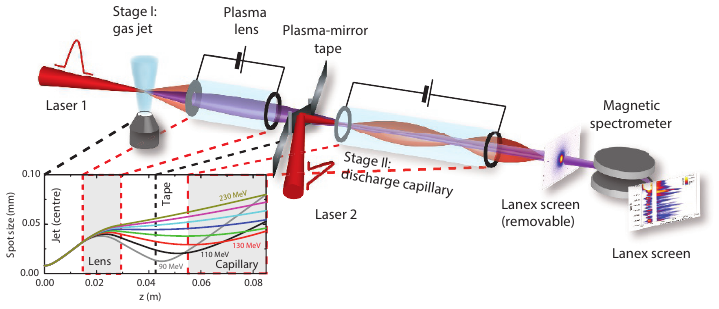}
    \caption{Schematic setup of a staging experiment using two independently laser-driven stages. An electron bunch injected in the first stage was focused using an active plasma lens. Since the energy spread of this bunch was large, the focus at the entry of the second stage was highly chromatic (see inset), resulting in significant emittance growth and consequently poor charge coupling (around 3.5\%). A plasma-mirror was used to couple in the second laser driver, which further contributed to emittance growth. Source: Ref.~\cite{Steinke2016} (reproduced with permission).}
    \label{fig:staging-experiment}
\end{figure}

Mirrors can be used to couple in and out laser drivers, often much more compactly than using dipoles. However, the laser intensity is too high close to the accelerator stage for using regular optical mirrors---the mirrors need to be placed several meters away (indeed the case in many single-stage laser-plasma-accelerator experiments). Instead, for compactness, \textit{plasma mirrors} have been proposed~\cite{Thaury2007}. Typically composed of a thin foil, the material is instantly ionized by the laser, which is then reflected by the resulting high-density plasma, while the particle bunch is allowed to pass straight through. This scheme was employed in the first successful demonstration of staging, performed at Lawrence Berkeley National Laboratory~\cite{Steinke2016} (see Fig.~\ref{fig:staging-experiment}). In this experiment, plastic tape was used as a plasma mirror to transversely couple in the second laser driver approximately 1~cm upstream of the second accelerating stage.

However, plasma mirrors can result in emittance growth for particle beams passing through them, both from Coulomb scattering (Section~\ref{sec:scattering}) as well as the so-called \textit{current filamentation instability}~\cite{Allen2012,Raj2020}, which can occur when the beam size is larger than the plasma skin depth, breaking the beam into many thin filaments. To avoid these issues, plasma mirrors must be made extremely thin---for instance by use of liquid-crystal films~\cite{Poole2016}, which can be made as thin as 10~nm. Emittance measurements of a laser-plasma-accelerated electron bunch has shown that the emittance growth from passing through such a liquid-crystal plasma mirror can be negligible (0.1~mm-mrad at GeV-level energies)~\cite{Zingale2021}.

\subsection{Chromaticity}
\label{sec:chromaticity}
Emittance growth from chromaticity is one of the biggest challenges of staging advanced accelerators~\cite{Antici2012,Migliorati2013}, and complicates the delivery of beams to applications such as free-electron lasers~\cite{Huang2012,Andre2018}. Two types of chromaticity are often discussed in the literature: (1) energy-dependent phase advance, and (2) energy-dependent focusing or Twiss parameters. Phase-advance chromaticity is very important in circular accelerators, and must be suppressed to avoid resonances. However, this is not normally a problem for single-pass linacs; the category within which most advanced accelerator schemes fall. In advanced accelerators, focusing chromaticity is more important due to small beta functions and strong focusing. This can clearly be seen in Fig.~\ref{fig:staging-experiment} (see inset), and explains why in this proof-of-principle experiment only about 3.5\% of the charge could be coupled and accelerated in the second stage~\cite{Steinke2016}.

\begin{figure}[b]
    \centering
    \includegraphics[width=0.55\linewidth]{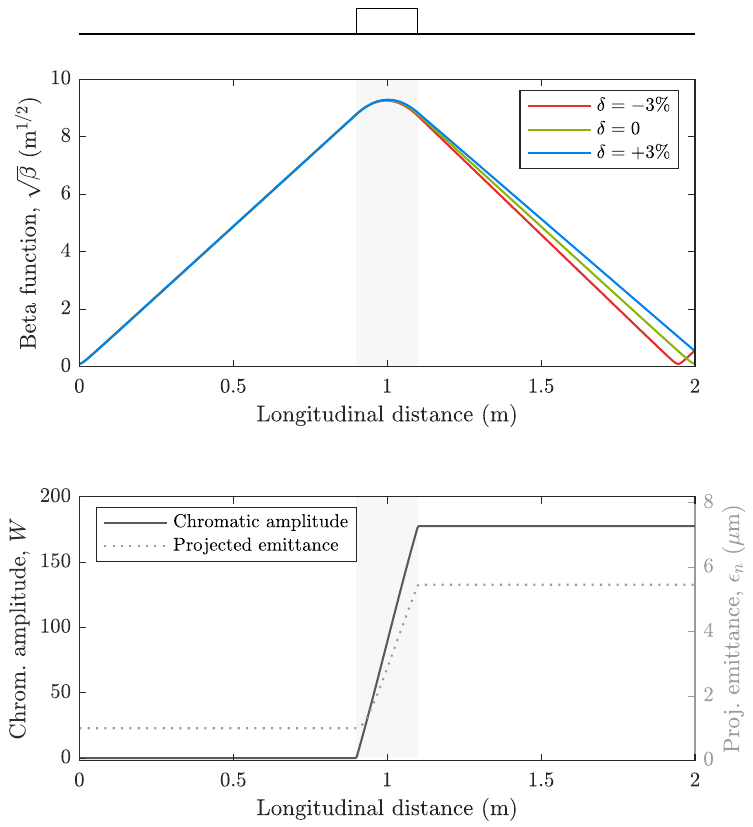}
    \caption{Chromaticity in a simplified staging setup with a single, radially focusing lens. Here, a beam with energy 10~GeV and energy spread 3\%~rms diverges from a matched beta function of approximately 10~mm (corresponding to a plasma density of $10^{16}$~cm$^{-3}$). The result is that the projected emittance grows by about 450\%. Source: Ref.~\cite{Lindstrom2021b} (CC BY 4.0).}
    \label{fig:chromaticity}
\end{figure}

Chromaticity is often quantified (to first order) in terms of the \textit{chromatic amplitude}
\begin{equation}
    W = \sqrt{\left(\frac{\partial\alpha_x}{\partial\delta}-\frac{\alpha_x}{\beta_x}\frac{\partial\beta_x}{\partial\delta}\right)^2 + \left(\frac{1}{\beta_x}\frac{\partial\beta_x}{\partial\delta}\right)^2},
\end{equation}
where $\delta$ is the relative energy offset; it is effectively a mismatch parameter for different energy slices. For a more complete description of chromaticity, one also needs to consider the chromatic phase, which oscillates at twice the rate of the usual phase advance. However, the chromatic amplitude is sufficient for calculating the relative projected (squared) emittance growth~\cite{Lindstrom2016b},
\begin{equation}
    \frac{\Delta(\epsilon^2)}{\epsilon_{\mathrm{init}}^2} = W^2 \sigma_{\delta}^2 + \mathcal{O}(\sigma_{\delta}^4).
\end{equation}
In beam optics with linear focusing, the chromatic amplitude grows by approximately $\Delta W = \beta_x/f$, where $f$ is the focal length---this implies that combined strong focusing (i.e., short focal length) and large beta functions result in large chromaticity and hence large emittance growth if left unmitigated. Staging, in its simplest form, can be considered as a single lens of focal length $f = L/2$, where $L$ is the distance from the accelerator to the lens (the distance between stages is $2L$), as illustrated in Fig.~\ref{fig:chromaticity}. In this case, the emittance growth from chromaticity in staging can be estimated to be
\begin{equation}
    \label{eq:staging-chromaticity}
    \frac{\Delta(\epsilon^2)}{\epsilon_{\mathrm{init}}^2} = \frac{4L^2}{\beta_m^2} \sigma_{\delta}^2
\end{equation}
to lowest order, given that $W \approx 2L/\beta_m$ since the beta function diverges in vacuum from $\beta_m \ll L$ to approximately $L^2/\beta_m$ at the position of the lens. The example in Fig.~\ref{fig:chromaticity} demonstrates that this emittance growth can be substantial for a typical plasma accelerator.

It should be noted that the emittance growth is only for the projected beam (i.e., there is no emittance growth within an energy slice). Only when the beam enters the following stage will the phase space be "scrambled" and the emittance growth made irreversible. If, however, the phase-space rotation could be reversed, the emittance could in principle be preserved. While this is impossible to do using linear optics, the chromaticity itself can be cancelled for a small range of energies (typically $\sigma_{\delta} \approx 1/W$) using a sufficient number of lenses; a technique known as \textit{apochromatic correction}~\cite{Montague1985,Lindstrom2016b}, often used in camera lenses.

To truly mitigate the effects of chromaticity, nonlinear optics is required. Normally, sextupole magnets are used in combination with dipoles, whereby particles of different energies are dispersed across the aperture of the sextupole, which has a different focal length at different transverse offsets. This technique enables focusing of a wide range of energies achromatically, and is therefore the go-to method for final focusing in linear colliders---a problem with fairly similar constraints to that of staging between advanced accelerators. A particularly potent concept is \textit{local chromaticity correction}~\cite{Raimondi2001}, where families of identical sextupoles are separated by 180 degrees of phase advance in order to cancel the deleterious effects of the higher-order geometric terms (i.e., $x^2$, $y^2$, and $xy$) from the nonlinear sextupole field.

Both apochromatic and nonlinear chromaticity correction can take up significant space between stages, which decreases the average accelerating gradient of a multistage accelerator. Minimizing the intrinsic chromaticity can reduce the need for corrective optics and possibly shrink the distance between stages. Equation~\ref{eq:staging-chromaticity} indicates three approaches: 
\begin{itemize}

    \item \textit{Lower energy spread.} Reducing the energy spread will reduce the emittance growth independently of the optics between the stages. However, this demands precise flattening of the wakefields~\cite{Tzoufras2008} and may not be compatible with the large energy-chirp required when employing BNS damping (Section~\ref{sec:beam-breakup-instability}).
    
    \item \textit{Larger matched beta functions.} Weaker in-situ focusing, either throughout the accelerator stage~\cite{Schroeder2013} or using ramps at the entry and exit can be used for increasing the matched beta function. However, ramps are not necessarily less chromatic---only \textit{adiabatic} ramps (where $\alpha_x \approx 0$ throughout) are near achromatic~\cite{Floettmann2014}. Unfortunately, adiabatic ramps can be fairly long, which can lead to emittance growth from scattering (Section~\ref{sec:scattering}) as well as poor energy efficiency from deceleration in low-density wakefields.
    
    \item \textit{Stronger focusing.} Shorter focal length beam-focusing devices will reduce the added chromatic amplitude because the beta function remains smaller throughout the focusing. Proposals include high-gradient permanent~\cite{Lim2005} and microelectromechanical~\cite{Harrison2015} quadrupole magnets, as well as plasma lenses (see Section~\ref{sec:plasma-lenses}). These advanced focusing devices not only reduce emittance growth from chromaticity, but also directly contributes to compactness, and will therefore likely be a key part of solving the staging problem.
    
\end{itemize}

\subsection{Plasma lensing}
\label{sec:plasma-lenses}
Comparing advanced accelerator optics, plasma lenses have two major advantages over permanent and microelectromechanical quadrupoles: (1) a potential for higher focusing gradients, and (2) azimuthally symmetric focusing, which removes the need for multiple lenses in order to capture diverging beams. This means plasma lensing can provide the shortest available focal length. However, plasma lenses also introduce new sources of emittance growth that need to be suppressed.

Two types of plasma lenses exist: \textit{passive plasma lenses}~\cite{Chen1987} are essentially short plasma accelerators using the very strong focusing fields of an ion-column, which can reach the equivalent of MT/m gradients as opposed to 1--100~T/m in conventional quadrupoles; and \textit{active plasma lenses}~\cite{Panofsky1950}, which focus charged particle beams using the magnetic field induced by a large (externally or "actively" driven) on-axis plasma current, reaching up to multi-kT/m gradients~\cite{vanTilborg2015,Sjobak2021}.

\subsubsection{Passive plasma lenses}

\begin{figure}[t]
    \centering
    \includegraphics[width=0.55\linewidth]{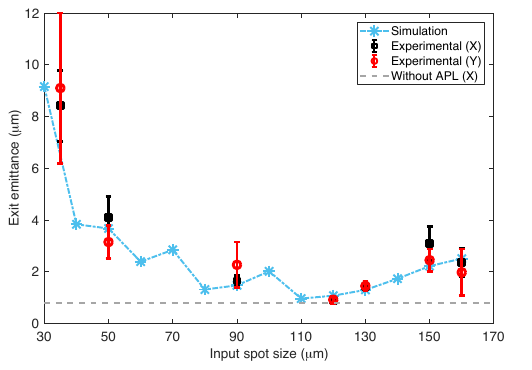}
    \caption{Emittance growth and preservation in a hydrogen-filled active plasma lens, in an experiment performed at INFN. For small beam sizes (left), the emittance grows due to a nonuniform wakefield excitation (passive plasma lensing), whereas for large beam sizes (right), the beam samples the nonlinear fields caused by the nonuniform radial temperature distribution. These two effects are both suppressed at one particular beam size (120~{\textmu}m rms) for the current profile used, resulting in emittance preservation. Source: Ref.~\cite{Pompili2018} (reproduced with permission).}
    \label{fig:active-passive-plasma-lens}
\end{figure}

Since passive plasma lenses are based on plasma wakefields, they will have many of the same sources of emittance as a plasma-accelerator stage (as discussed in Section~\ref{sec:emittance-growth-in-stage}). Since the lens is short, however, instabilities have no time to form, and since the beam size will typically be larger, ion motion is unlikely to occur. If the bunch interacts with its own wake, it will experience transverse beam loading (Section~\ref{sec:transverse-beam-loading})---that is, longitudinally and transversely nonuniform focusing. This can be avoided for electron bunches by using a separate driver to drive a nonlinear wake \cite{Thaury2015}, in which the focusing can be fully uniform. However, this also requires in- and out-coupling for the driver, which can cause its own emittance growth (Section~\ref{sec:driver-in-and-out-coupling}). Lastly, Coulomb scattering (Section~\ref{sec:scattering}) can also occur, but is usually negligible due to the low plasma densities and short propagation distances required in a passive plasma lens.

\subsubsection{Active plasma lenses}

Active plasma lenses avoid the problems of longitudinal nonuniformity due to the field being imposed externally---at least in principle. If the focused beams have sufficiently high currents or small beam sizes, wakefields are driven~\cite{Lindstrom2018c}, effectively turning the active plasma lens into a passive plasma lens, with all the above-mentioned sources of emittance growth. Moreover, active plasma lenses can suffer from intrinsic aberrations due to nonuniform radial temperature distributions~\cite{Bobrova2001,vanTilborg2017}. This is caused by plasma cooling near the walls, leading to increased resistance and therefore lower current density at larger radii, which again results in a nonlinear field and therefore emittance growth. Recently, it was found that the aberration does not occur in heavier gases such as argon (as opposed to lighter gases like hydrogen and helium)~\cite{Lindstrom2018b}. However, increasing the atomic number also causes a dramatic increase in emittance growth from scattering, which is already more severe in an active plasma lens due to the higher plasma density and longer lengths required. Emittance preservation has been experimentally demonstrated in active plasma lenses~\cite{Lindstrom2018b,Pompili2018} (see Fig.~\ref{fig:active-passive-plasma-lens}), but it is still unclear whether these results can be scaled to a full-length advanced accelerator.

\section{Conclusions}

In summary, emittance preservation in an advanced accelerator is a delicate balancing act. Many sources of emittance growth exist both inside and outside the accelerator stages. Inside a stage, emittance can grow from instabilities due to transverse wakefields, slice decoherence in mismatched or misaligned beam of finite energy spread, nonlinear focusing from ion motion or transverse beam loading, as well as beam--gas scattering. Between stages, emittance growth can occur in plasma mirrors, by leaking dispersion from dipole chicanes, from chromatic focusing of highly diverging beams, or because of new problems from new types of optics such as plasma lenses. Mitigation strategies for each of these effects exist, but form a complex web of interconnections, often with conflicting scalings. Ultimately, this means that the effects must be balanced such that suppressing one source of emittance growth does not blow up another.

The key to finding this balance, and thereby enabling \textit{usable} advanced accelerators, will likely be a combination of accurate numerical modelling of all known effects, precise benchmarking from experiments, as well as efficient methods of optimizing systems with many parameters. Somewhere in this vast and largely unexplored parameter space, surely, regions will exist that allow sustained high-gradient acceleration while preserving emittance and other beam qualities.


\end{document}